\documentclass{article}

\usepackage[normalem]{ulem} 
\usepackage{amsmath,amsfonts}
\usepackage{hyperref}

\usepackage{graphicx}

\usepackage{siunitx}
\usepackage[capitalise]{cleveref}

\crefname{equation}{}{} 
\Crefname{equation}{Eq.}{Eqs.}

\usepackage{xspace}

\usepackage{amsthm}
\theoremstyle{remark}
\newtheorem*{remark}{Remark}

\usepackage{authblk}

\usepackage{natbib}
\setcitestyle{authoryear,round}

\renewcommand{\d}{\partial}

\newcommand{\permeability}{\ensuremath{K}\xspace}
\newcommand{\porosity}{\ensuremath{\phi}\xspace}
\newcommand{\velocity}{\ensuremath{u}\xspace}

\newcommand{\fg}{\ensuremath{f_\text{g}}\xspace}
\newcommand{\fgm}{\ensuremath{f_\text{g}^-}\xspace}

\newcommand{\fw}{\ensuremath{f_\text{w}}\xspace}

\newcommand{\fwm}{\ensuremath{f_\text{w}^-}\xspace}

\newcommand{\Kc}{\ensuremath{K_\text{c}}\xspace}

\newcommand{\krg}{\ensuremath{k_\text{rg}}\xspace}

\newcommand{\krw}{\ensuremath{k_\text{rw}}\xspace}

\let\lg\relax

\newcommand{\lg}{\ensuremath{\lambda_\text{g}}\xspace}
\newcommand{\li}{\ensuremath{\lambda_\text{i}}\xspace}

\newcommand{\lw}{\ensuremath{\lambda_\text{w}}\xspace}

\newcommand{\mug}{\ensuremath{\mu_\text{g}}\xspace}

\newcommand{\muw}{\ensuremath{\mu_{\text{w}}}\xspace}

\newcommand{\nd}{\ensuremath{n_\text{D}}\xspace}

\newcommand{\ndLE}{\ensuremath{n_\text{D}^{\text{LE}}}\xspace}
\newcommand{\ndm}{\ensuremath{n_\text{D}^-}\xspace}
\newcommand{\ndp}{\ensuremath{n_\text{D}^+}\xspace}
\newcommand{\ndpm}{\ensuremath{n_\text{D}^\pm}\xspace}

\newcommand{\nf}{\ensuremath{n_\text{f}}\xspace}
\newcommand{\nmax}{\ensuremath{n_{\max}}\xspace}

\newcommand{\pc}{\ensuremath{P_\text{c}}\xspace}

\newcommand{\rc}{\ensuremath{r_\text{c}}\xspace}
\newcommand{\rg}{\ensuremath{r_\text{g}}\xspace}

\newcommand{\RR}{\ensuremath{\mathbb{R}}\xspace}

\newcommand{\Sg}{\ensuremath{S_\text{g}}\xspace}

\newcommand{\Sgr}{\ensuremath{S_\text{gr}}\xspace}

\newcommand{\Swc}{\ensuremath{S_\text{wc}}\xspace}

\newcommand{\Swe}{\ensuremath{S_\text{we}}\xspace}

\newcommand{\Sw}{\ensuremath{S_\text{w}}\xspace}

\newcommand{\Swm}{\ensuremath{S_\text{w}^-}\xspace}
\newcommand{\Swp}{\ensuremath{S_\text{w}^+}\xspace}
\newcommand{\Swpm}{\ensuremath{S_\text{w}^\pm}\xspace}

\newcommand{\ug}{\ensuremath{\velocity_\text{g}}\xspace}

\newcommand{\uw}{\ensuremath{\velocity_\text{w}}\xspace}

\newcommand{\vg}{\ensuremath{v_\text{g}}\xspace}
\newcommand{\vw}{\ensuremath{v_\text{w}}\xspace}

\newcommand{\dpc}[0]{\frac{\d \pc}{\d S_w}}

\newcommand{\pressure}[0]{\ensuremath{P}\xspace}

\newcommand{\tildeSw}{\ensuremath{\tilde{S}_\text{w}}\xspace}
\newcommand{\tildenf}{\ensuremath{\tilde{n}_\text{D}}\xspace}
\newcommand{\tildenD}{\ensuremath{\tilde{n}_\text{D}}\xspace}
\newcommand{\tildemuf}{\ensuremath{\tilde{\mu}_\text{g}}\xspace}
\newcommand{\tildeuw}{\ensuremath{\tilde{u}_\text{w}}\xspace}

\newcommand{\kri}{\ensuremath{k_\text{ri}}\xspace}
\newcommand{\mui}{\ensuremath{\mu_\text{i}}\xspace}

\title{Traveling wave solutions for non-Newtonian foam flow in porous media}
\author[1,2]{Weslley da Silva Pereira, \href{mailto:weslley.pereira@ucdenver.edu}{weslley.pereira@ucdenver.edu}}
\author[1]{Grigori Chapiro, \href{mailto:grigori@ice.ufjf.br}{grigori@ice.ufjf.br}}
\affil[1]{Department of Mathematics, Federal University of Juiz de Fora, Rua Jos\'e Louren\c{c}o Kelmer, s/n - S\~ao Pedro, Juiz de Fora, 36036900, MG, BR}
\affil[2]{Department of Mathematical and Statistical Sciences, University of Colorado Denver, 1201 Larimer Street, Denver, 80204, CO, US}

\begin{document}

\maketitle

\abstract{The injection and in-situ generation of foam in porous media successfully control gas mobility and improve the fluids' sweep efficiency inside porous media.
Mathematical models describing this problem use two phases, foamed gas and fluid, and usually have a term for foam generation and destruction.
Moreover, the non-Newtonian foam behavior is frequently modeled using the Hirasaki and Lawson's formula for foamed gas viscosity.
In this paper, we detail how the traveling wave analysis can be used to estimate the propagation profiles and velocity for a range of non-Newtonian foam models in porous media at constant total superficial flow velocity.
We reformulate Hirasaki and Lawson's formula in an explicit form allowing us to find traveling wave solutions for the non-Newtonian Linear Kinetic model. Comparing the solution with the one for the Newtonian version, allows us to analyze qualitatively and quantitatively the rheology of the foam flow in porous media.}

\subsection*{Funding}
This research was carried out in association with the ongoing R\&D project registered as ANP 20715-9, ``Modelagem matem\'atica e computacional de inje\c{c}\~ao de espuma usada em recupera\c{c}\~ao avan\c{c}ada de petr\'oleo" (Universidade Federal de Juiz de Fora (UFJF) / Shell Brasil / ANP) - Mathematical and computational modeling of foam injection as an enhanced oil recovery technique applied to Brazil pre-salt reservoirs, sponsored by Shell Brasil under the ANP R\&D levy as ``Compromisso de Investimentos com Pesquisa e Desenvolvimento". This project is carried out in partnership with Petrobras. The author G. Chapiro was supported in part by CNPq grant 303245/2019-0.

\section{Introduction}\label{sec1}

One technique to control gas mobility and improve fluid sweep efficiency in a porous medium consists of injecting foaming agents, e.g., surfactants and nanoparticles, in aqueous solutions that create bubbles in the gas phase.
Several studies point out the favorable application of this method in enhanced oil recovery, e.g., \citep{Casteel1988,rossen2014foamIOR}, acid diversion during matrix stimulation, e.g., \citep{Behenna1995}, and contaminated aquifer remediation, e.g., \citep{Burman1986}.

The behavior of foam is frequently reported as shear-thinning \citep{Ferno2016,Heller1987,Marsden1966,Zitha2006New}, which means that its apparent viscosity decreases with increased shear stress. \cite{Hirasaki1985} derived a shear-thinning expression for the apparent viscosity of foam flowing in capillary tubes and validated it through laboratory experiments.
This non-Newtonian formula is consistent with the classical result of \cite{Bretherton1961} using isolated bubbles and has application in several foam propagation models, e.g., \citep{Izadi2019,Kam2008,Kovscek1995,Simjoo2015}.
Hirasaki and Lawson's expression implicitly defines the apparent gas viscosity as it depends on the gas velocity, and the gas velocity depends on the viscosity.
Other works report that foam behaves as a Newtonian fluid for particular regimes. For example, foam flow can be classified as Newtonian under nearly-constant capillary pressure during steady radial flow \citep{R.Rossen1991}. Following this idea, some models consider foam flow to be a Newtonian fluid \citep{Ashoori2011a,Lozano2021} and present results compatible with experimental data \citep{Zavala2021}.
Hirasaki and Lawson's formula is so important that numerous papers investigate the foam flow in porous media only focusing on the apparent viscosity (or, equivalently, the mobility reduction factor).

The studies reported above motivate the current discussion on when to consider non-Newtonian behavior of the foam flow. \cite{Vassenden1998} present and validate a model for the transition of foam flow behavior from Newtonian to shear-thinning according to the gas flow rate. \cite{Alvarez2001a} conclude the foam behavior is shear-thinning in the low-quality regime and shear-thickening in high-quality regimes. \cite{R.Rossen1991} highlights the change from Newtonian to non-Newtonian behavior of uniform texture foam for changing capillary pressure.

One may find several predictive mathematical models in the literature aiming to represent foam propagation inside porous media.
Local Equilibrium (LE) foam models use algebraic expressions to compute the foam texture. However, LE models can be inadequate when strong foam generation is not certain \citep{Kam2007}.
Mechanistic foam models, introduced by \cite{Falls1988,Patzek1988}, describe the foam texture dynamically using partial differential equations (PDE). They have the unique potential to describe both transient and stationary conditions for foam flow as they track the foam texture in space and time \citep{Eide2020}.
Many works report mechanistic models to match laboratory experiments successfully, e.g., in \citep{Kovscek1995,Kam2008,Simjoo2015}.
Besides that, some authors prefer these models when extracting analytical estimates of the foam flow, e.g., \citep{Almajid2019,Eide2020,Simjoo2015}.
Moreover, several works suggest that, after a transient stage, foam travels inside the porous media under invariant water saturation profiles, e.g., \citep{Vries1990,Ettinger1992}. One strategy to obtain solution profiles, employed in \citep{Izadi2019,Kam2008,Simjoo2015}, consists of disregarding the capillary pressure gradient in each phase's superficial velocity and applying the classical method of characteristics. Another strategy consists of seeking traveling waves solutions, which are profiles invariant to translation in space \citep{Volpert2000}. They are especially suited for analyzing nonlinear wave propagation problems at constant speed. \cite{Ashoori2011a,Lozano2021} used this procedure to find traveling wave solutions for Newtonian foam models at fixed total superficial velocity. \cite{Ashoori2011a} also find traveling waves for the model from \cite{Kam2008} which does not consider capillary pressure derivatives.

The present work shows how to compute traveling wave solutions that connect two equilibrium states of a generic mechanistic non-Newtonian foam model that consider capillary pressure derivatives.
We assume the total superficial velocity is fixed, which is the case of predominant rectilinear flow.
The non-Newtonian foamed gas viscosity is modeled by the Hirasaki and Lawson's formula. We rewrite this formula in terms of water saturation, foam texture and total superficial velocity to avoid the direct dependency on the gas superficial velocity.
The traveling wave profiles serve as basis to compare a linear kinetic foam model using Newtonian and non-Newtonian gas viscosity expressions.

We organize this paper as follows.
\Cref{sec:model} presents the general population-based foam model. \Cref{sec:newtonianVSshearthinningMobilities} shows the Hirasaki and Lawson apparent gas viscosity and its reformulated version. \Cref{sec:travWaves} presents the procedure for obtaining traveling wave solutions for the non-Newtonian foam flow model. \Cref{sec:MRFandfw} applies the procedure to a linear kinetics model, and use the traveling wave profiles to compare Newtonian and non-Newtonian models. \Cref{sec:conclusions} presents the discussions and conclusions.

\section{The population-balance foam model}
\label{sec:model}
Consider the two-phase flow of gas-water solution with foaming agents in a porous medium, where the flow occurs only in one direction, and all fluids are incompressible. The last assumption is undesirable but necessary for the mathematical analysis performed in this paper.
The process is modeled by the mass balance of each phase $i$ as 
\begin{align}\label{eq:massConsLaw}
	\frac{\d(\porosity\,S_i)}{\d t} + \frac{\d u_i}{\d x} = 0\,, \qquad i \in \{w,g\}\,,
\end{align}
where
	$t$ is the time variable,
	$x$ is the space variable,
	$\phi$ is the porosity,
	$S_i$ and $u_i$ are the saturation and superficial velocity of the phase $i$.
The subscript `$w$' represents the aqueous phase and `$g$' represents the gas phase.
The phase superficial velocities satisfy the generalized Darcy's laws \citep{Chen1997a}
\begin{align}
	\label{eq:partialuw}
	\uw &= \fw\,\left(u + \lg\,\permeability\,\frac{\d \pc}{\d x}\right)\,,\qquad
	\ug = \fg\,\left(u - \lw\,\permeability\,\frac{\d \pc}{\d x}\right)\,,
\end{align}
with fractional fluxes $f_i$ and phase mobilities $\li$ given by
\begin{align}
	f_i &:= \frac{\li}{\lw+\lg}\,,\qquad 
	\li := \frac{\kri}{\mui}\,,
	\qquad i \in \{w,g\},
	\label{eq:phaseMobilities}
\end{align}
where
	$\permeability$ is the permeability,
	$u = u_w + u_g$ is the (constant) total superficial velocity,
	$\pc$ is the capillary pressure,
	and for each phase $i$,
	$\kri$ is the relative permeability, and
	$\mui$ is the apparent phase viscosity.
The balance of foam texture $\nf$ in the gas phase is
\begin{equation}\label{eq:modelFoam}
	\frac{\d(\porosity\,\Sg\,\nf)}{\d t} + \frac{\d(\nf\,\ug)}{\d x} = \porosity\,\Sg\,R\,,
\end{equation}
where $R := \rg - \rc$, and $\rg$ and $\rc$ are the rates of generation and coalescence of foam, respectively.
It is useful to define the dimensionless foam texture ($\nd$) as follows
\begin{equation}\label{eq:nDFoam}
	\nd := \frac{\nf}{\nmax}\,,
\end{equation}
where $\nmax$ is the reference foam texture.

Some auxiliary physical quantities are equally important in this work.
We define the effective water saturation ($\Swe$), with  values in [0,1], as
\begin{align}
	\Swe := \frac{\Sw-\Swc}{1-\Swc-\Sgr}\,,
\end{align}
where
	$\Swc$ is the connate water saturation, and
	$\Sgr$ is the residual gas saturation.
The total apparent viscosity ($\mu_{app}$) is
\begin{align}\label{eq:muApp}
	\mu_{app} := \frac1{\lw + \lg}\,,
\end{align}
and the mobility reduction factor (MRF) is
\begin{align}\label{eq:MRF}
	MRF := \frac{\mug}{\mug^0}\,.
\end{align}
To avoid problems in the nomenclature, we emphasize using the word ``total'' when referring to the apparent viscosity $\mu_{app}$.
Note that the apparent total viscosity represents the equivalent viscosity of the fluid system in the following Darcy's law
\begin{align}\label{eq:pApp}
	u = -\frac{\permeability}{\mu_{app}} \nabla \pressure\,,
\end{align}
where $P$ is the global pressure \citep{Antontsev1972}.
Finally, the phase-$i$ interstitial velocity is
\begin{align}\label{eq:vi}
	v_i := \frac{\vert\velocity_i\vert}{\porosity\,S_i}\,,
	\qquad i \in \{w,g\}.
\end{align}

It is worth mentioning that several models exist that state $\pc$ depending exclusively on the fluid phase saturation, and many of them are based on the classical works of
\cite{Corey1954,CoreyBrooks1966capillary,vanGenuchten1980capillary}.
The relative permeabilities $\krw$ and $\krg$ are either derived directly from the capillary pressure or fitted experimentally (see \citep{Li2006} and references therein).
Foam generation and coalescence expressions $\rg$ and $\rc$ are either inspired on microscopic mechanisms \citep{Kam2008,Kovscek1995}, based on macroscopic observations \citep{Ashoori2011a,Simjoo2015}, or fitted using laboratory experiments \citep{Thorat2016foam}.
We postpone the definitions of the expressions $\krw(\Sw)$, $\krg(\Sw)$, $\rg(\Sw,\nd,\uw)$, $\rc(\Sw,\nd,\uw)$, and $\pc(\Sw)$ to \cref{sec:MRFandfw}.
\Cref{sec:newtonianVSshearthinningMobilities} presents the formulas for the apparent gas viscosity ($\mug$).

Some mechanistic (population-based) models use an expression for LE foam texture, $\ndLE$, that naturally arises from the restriction $R=0$. For example:
\begin{itemize}
    \item The model from \cite{Simjoo2015} uses a constant value $\ndLE$, which is compatible with laboratory experiments.
    \item In \citep{Ashoori2011a}, $R=0$ if and only if $\nd = \ndLE(\Sw) = \tanh (400\,(\Sw-0.37))$, which approximates the Heaviside step function.
    \item In \citep{Chen2010}, $\ndLE$ is the only real solution of
\begin{align}\label{eq:Chen2010ndLE}
(\ndLE)^3 + \frac{k_{-1}\vg^{2/3}}{k_1^0\vw}\ndLE - 1 = 0\,,
\end{align}
where $k_{-1}$ and $k_1^0$ are positive coefficients possibly depending on $\Sw$.
\end{itemize}
In the examples above, $0 \le \ndLE \le 1$.
The expression $\ndLE$ from \cite{Kovscek1995} can be also written as a function of $\Sw$, $\vw$ and $\vg$ but it is not constrained to the interval $[0,1]$. In \citep{Kam2008}, $\ndLE$ is also not constrained to the interval $[0,1]$ and it is a function of $\Sw$ and $\nabla \pressure$ (See \citep{Ashoori2011a} for a traveling wave analysis of the model in \citep{Kam2008}).

\begin{remark}
Some models make a distinction between trapped and flowing foam phases, e.g., \citep{Izadi2019,Kovscek1995,Kharabaf1998}. Other studies report the importance of understanding the gas trapping mechanism to model foam flow in porous media \citep{Almajid2019,Jones2018TrappedGas}.
\cite{Simjoo2015} observe no trapping, arguing that foam could be moving slowly.
On top of that, it is still not clear how to model trapped foam fraction, although some models try to explain it \citep{Kovscek1995,Cohen1997a,Tang2006}.
The analysis presented in this paper may be applied to some trapped foam models, for instance, the model from \cite{Kovscek1995} and its derivatives.
\end{remark}

\section{The Hirasaki and Lawson's gas viscosity}
\label{sec:newtonianVSshearthinningMobilities}

The apparent gas viscosity in the presence of foam is frequently modeled by the Hirasaki and Lawson's formula
\begin{equation}\label{eq:viscosityHirasaki}
	\mug = \mug^0 + \alpha\,\frac{\nf}{\vg^{1/3}}\,,
\end{equation}
when foam is considered a shear-thinning fluid, where $\mug^0$ is the gas viscosity in the absence of foam.
The proportionality constant $\alpha$ depends on many factors, including the pore structure, the liquid viscosity, and the gas-liquid surface tension \citep{Hirasaki1985}. Due to the inherent difficulty to obtain $\alpha$, many works use it as a fitting parameter, e.g., \citep{Chen2010,Eide2020,Kam2008,Kovscek1995,Simjoo2015,Thorat2016foam}.
Since $\vg$ itself depends on $\mug$ (see \cref{eq:partialuw,eq:phaseMobilities,eq:vi}), \Cref{eq:viscosityHirasaki} defines $\mug$ implicitly, which hinders the analytical and numerical analysis of the model.

The first contribution of this work is to rewrite \cref{eq:viscosityHirasaki} eliminating the direct dependency on $\vg$. To do so, we replace \eqref{eq:partialuw}, \eqref{eq:phaseMobilities}, and \eqref{eq:vi} into \eqref{eq:viscosityHirasaki} obtaining a cubic equation in the variable $X = (\krg+\lw\,\mug)^{1/3}$.
The resulting expression admits a unique positive root $X$, which can be written as
\begin{align}\label{eq:newViscosityTotalPressure}
	\mug &= \mug^0 + \frac{3\,A}{\lambda_w}\,\left(\sqrt[3]{ B + \sqrt{B^2 - A^3} } + \sqrt[3]{ B - \sqrt{B^2 - A^3} }\right)\,,
\end{align}
where
\begin{align}
	A &:= 
	\frac{\alpha\,\nf\,\lw}{3}\,\left(\frac{\porosity\,(1-\Sw)}{\krg\,
	\left \vert u - \lw\,\permeability \, \partial_x\pc \right \vert  }\right)^{1/3} 
	\quad\text{ and }\quad
	B := \frac{\krg+\lw\,\mug^0}{2}\,.
\end{align}
Under LE conditions, the capillary pressure is constant. Therefore, term $A$ in \eqref{eq:newViscosityTotalPressure} becomes
\begin{align} 
	A &:= \frac{\alpha\,\nmax\,\ndLE\,\lw}{3}\,\left(\frac{\porosity\,(1-\Sw)}{\krg\,\vert u\vert }\right)^{1/3}.
\end{align}
Since $\lw$ and $\krg$ are functions of $\Sw$, and $\mug^0$, $\alpha$, $\nmax$, and $\phi$ are constant parameters, we identify the apparent gas viscosity under LE conditions as the function $\mug^{LE}(\Sw,\ndLE,\velocity)$.

\section{Traveling wave solutions}
\label{sec:travWaves}

\cite{Ashoori2011a,Lozano2021} investigated traveling wave solutions for a linear kinetic Newtonian foam model, where the apparent gas viscosity ($\mug$) depends only on the foam texture ($\nd$). Below, we present the analysis for the generic non-Newtonian model described in \cref{sec:model} with apparent viscosity formula detailed in \cref{sec:newtonianVSshearthinningMobilities}.

\subsection{Traveling wave formulation}
\label{sec:travelWavesODE}

The traveling wave solutions of a system of evolutionary PDEs are the solutions invariant to translation in space (see \citep{Volpert2000} and references therein). Mathematically speaking, they depend on a traveling variable $\xi := x-vt$, where $v$ is the constant wave velocity.
For the system \eqref{eq:massConsLaw}, \eqref{eq:partialuw} and \eqref{eq:modelFoam} with step function initial conditions 
\begin{equation}
(\Sw,\nd) = \left\{ 
\begin{array}{cc}
(\Swm, \ndm),         & x=0 \\[5pt]
(\Swp, \ndp),        & x>0
\end{array}
\right.
\end{equation}
the traveling wave solution is represented by bounded and differentiable functions $\tildeSw, \tildenf : \RR \rightarrow \RR$  satisfying $(\tildeSw,\tildenf)(x-vt) = (\Sw,\nd)(x,t)$ for a some $v \in \RR$, all $x \in (0,L)$, $L>0$, and all $t \in (0,+\infty)$. Moreover, the wave front $(\tildeSw,\tildenf)$ must satisfy the boundary conditions \citep{Volpert2000}:
\begin{equation}
\label{eq:travelingODElimits}
	\lim_{\xi\rightarrow \pm\infty} (\tildeSw,\tildenf)(\xi) = (\Swpm, \ndpm) \in \RR^2\,,\quad
	\text{and}\quad
	\lim_{\xi\rightarrow \pm\infty} d_\xi (\tildeSw,\tildenf)(\xi) =
	0\,.
\end{equation}
The (left) state $(\Sw^-,\nd^-)$ corresponds to the inflow boundary conditions, and the (right) state $(\Sw^+,\nd^+)$ corresponds to the initial conditions for the original PDE.

Starting from \eqref{eq:massConsLaw}, \eqref{eq:partialuw} and \eqref{eq:modelFoam}, one may verify that the wave front $(\tildeSw,\tildenf)$ is a solution of the following system of ordinary differential equations (ODEs):
\begin{equation}
\label{eq:travelingODE}
    \left\{
    \begin{aligned}
        &d_\xi (-v\,\porosity\,\tildeSw + \tildeuw)
        = 0\,,\\
        &d_\xi \left[\tildenf\,\nmax\,(-v\,\porosity\,(1-\tildeSw) + (u - \tildeuw))\right]
        = \porosity\,(1-\tildeSw)\,\tilde{R},
    \end{aligned}
    \right.
\end{equation}
for all $\xi \in \RR$, where $\tilde{R} := R\left(\,\tildeSw\,,\;\tildenf\,,\;\tildeuw\,\right)$,
\begin{align}
\label{eq:travelingODEtildeu}
	\tildeuw
        = \frac{\lw(\tildeSw)}{\lw(\tildeSw)+\frac{\krg(\tildeSw)}{\tildemuf}}\,\left(u + \frac{\krg(\tildeSw)}{\tildemuf}\,\permeability\,\dpc(\tildeSw)\,\frac{\d\tildeSw}{\d \xi} \right)\,,
\end{align}
and $\tildemuf$ is given below in \cref{eq:viscosityHirasakiTravelingWave}.
The solution $(\tildeSw,\tildenf)$ of \eqref{eq:travelingODE} satisfying  restrictions \eqref{eq:travelingODElimits} is known as traveling wave solution.
We can use the limits in \eqref{eq:travelingODElimits} to conclude that there exists $\ug^+ := u - \lim_{\xi\rightarrow +\infty} \tildeuw$ such that 
\begin{align}\label{eq:ugPlus}
-v\,\porosity\,(1-\tildeSw) + (u-\tildeuw) = -v\,\porosity\,(1-\Swp) + \ug^+
\end{align}
for all $\xi \in \RR$.
Thus, we can rewrite \crefrange{eq:travelingODE}{eq:travelingODEtildeu} as follows
\begin{align}
\label{eq:travelingODE2}
	\left\{
    \begin{aligned}
        &\frac{-\lw\,\krg}{\krg+\lw\,\tildemuf}\,\permeability\,\dpc\, d_\xi \tildeSw
        = \left[\ug^+ - \frac{\krg\,u}{\krg+\lw\,\tildemuf}\right] - v\,\porosity\,(\tildeSw-\Swp)\,,\\
        &\nmax\,(\ug^+-v\,\porosity\,(1-\Swp))\, d_\xi \tildenf
        = \porosity\,(1-\tildeSw)\,\tilde{R}\,,
    \end{aligned}
    \right.
\end{align}

Note that \cref{eq:ugPlus} also implies that, for all $\xi \in \RR$,
\begin{equation}
\label{eq:travelingVg}
	\widetilde{\vg}
	:= \frac{\vert u-\tildeuw\vert }{\porosity\,(1-\tildeSw)}
	= \left\vert v\frac{\left(\Swp + \frac{\ug^+}{\porosity v}\right) - \tildeSw}{1-\tildeSw}\right\vert \,.
\end{equation}
Then, one may replace \eqref{eq:travelingVg} into \eqref{eq:viscosityHirasaki} to obtain
\begin{equation}\label{eq:viscosityHirasakiTravelingWave}
	\tildemuf = \mug^0 \left(1 + \frac{\alpha\,\nmax}{\mug^0\,\vert v\vert^{\frac13}}\,\left(
		\frac{1-\tildeSw}{\big\vert \Sw^{crit} - \tildeSw\big\vert }
	\right)^{\frac13}\,\tildenf\right)\,,
\end{equation}
where $\Sw^{crit} := \Swp+\ug^+/(v\,\phi)$.
Equation \eqref{eq:viscosityHirasakiTravelingWave} is an alternative version of \eqref{eq:viscosityHirasaki} valid only for the traveling wave profile.

Let us summarize these calculations. The traveling wave solution of the problem \eqref{eq:massConsLaw}-\eqref{eq:modelFoam} in the domain $(0,L) \times (0,+\infty)$, with initial condition $(\Sw,\nd)(x,0) = (\Swp,\ndp)$ and boundary condition $(\Sw,\nd)(0,t) = (\Swm,\ndm)$ is given by the solution of \cref{eq:travelingODE2} which satisfies limit conditions \eqref{eq:travelingODElimits}. We discuss how to obtain these limit conditions in Subsection \ref{sec:equilibria}.

The existence of the solution of \eqref{eq:travelingODE2} is not trivial. \cite{Lozano2021} investigate the existence of solutions for \cref{eq:travelingODE2} using Newtonian foam flow and particular choice of $R = \rg - \rc$. They show that such solutions exist in specific parameter regions, depending on the model and the limit states.
We do not perform such deep analysis here; however, we give evidence that the solutions exist for the examples from \cref{sec:MRFandfw}.

\subsection{Traveling wave velocity and equilibria}
\label{sec:equilibria}

The usual procedure to determine the traveling wave velocity and equilibria is to apply limits \cref{eq:travelingODElimits} to the ODE system \cref{eq:travelingODE2}.
Let us assume $\krw(\Swpm)$, $\krg(\Swpm)$, and $d_{\Sw}\pc(\Swpm)$ are real numbers and that $\krg(\Swpm) > 0$.
As $\lim_{\xi\rightarrow \pm\infty} d_\xi \tildeSw (\xi) = 0$ from \eqref{eq:travelingODElimits}, it follows that 
$\lim_{\xi\rightarrow \pm\infty} d_x\pc (\xi) = 0$.
Therefore, $\lim_{\xi\rightarrow \pm\infty} \tildemuf (\xi) = \mug^{LE}(\Swpm,\nd^\pm,\velocity)$.
Moreover, we use the fractional flux definition in \cref{eq:phaseMobilities}, and apply the limits in \cref{eq:travelingODElimits} to \cref{eq:travelingODE2,eq:travelingODEtildeu} to obtain the algebraic system of nonlinear equations
\begin{align}
\label{eq:travelingODElimitSystem_fgm}
    		&\frac{\krg(\Swm)}{\krg(\Swm)+\lw(\Swm)\,\mug^{LE}(\Swm,\nd^-,\velocity)}
    		= \fgm\,,\\
\label{eq:travelingODElimitSystem_ugp}
        &\frac{\krg(\Swp)}{\krg(\Swp)+\lw(\Swp)\,\mug^{LE}(\Swp,\nd^+,\velocity)}
        = \frac{\ug^+}{u}\,,\\
\label{eq:travelingODElimitSystem_v}
        &\left(\ug^+ - \fgm\,u\right) - v\,\porosity\,(\Swm-\Swp)
        = 0\,,\\
\label{eq:travelingODElimitSystem_rgrcm}
        &R\left(\,\Swm\,,\;\ndm\,,\;(u-\ug^+) + v\,\porosity\,(\Swm-\Swp)\,\right)
        = 0\,,\\
\label{eq:travelingODElimitSystem_rgrcp}
        &R\left(\,\Swp\,,\;\ndp\,,\;u-\ug^+\,\right)
        = 0\,.
\end{align}
Most laboratory and field scale simulations use homogeneous initial saturation profiles, and measure the injection gas fractional flow to predict foam propagation. Therefore, we consider $\velocity$, $\Swp$ and $\fgm$ to be the known values and  $(\Swm,\ndm,\ndp,\ug^+,v)$ as the main unknowns in \crefrange{eq:travelingODElimitSystem_fgm}{eq:travelingODElimitSystem_rgrcp}.

As discussed in \Cref{sec:model}, there is a particular case where the restrictions $R=0$ and $\partial_x\pc = 0$ result in a LE foam texture $\nd = \ndLE(\Sw,\uw,\velocity)$.
The models from \citep{Ashoori2011a,Chen2010,Kam2008,Kovscek1995,Simjoo2015}, for instance, allow for such an expression of local-equilibrium foam texture.
For these models, one may use the following strategy to solve \crefrange{eq:travelingODElimitSystem_fgm}{eq:travelingODElimitSystem_rgrcp} given $\velocity$, $\Swp$ and $\fgm$:
\begin{enumerate}
\item Solve \cref{eq:travelingODElimitSystem_fgm} for $\Swm$ using $\nd^- = \ndLE(\Swm,(1-\fg^-)\,\velocity,\velocity)$;
\item Solve \cref{eq:travelingODElimitSystem_ugp} for $\ug^+$ using $\nd^+ = \ndLE(\Swp,\velocity-\ug^+,\velocity)$;
\item Solve \cref{eq:travelingODElimitSystem_v} for $v$.
\end{enumerate}
This algorithm involves the solution of two uncoupled scalar nonlinear equations, which can be done either analytically or applying a classical numerical procedure, see \citep{Mathews1992}.
After obtaining all constant parameters, one must solve the ODE system \eqref{eq:travelingODE2} with limit conditions in \cref{eq:travelingODElimits} using usual ODE solvers for nonlinear systems to obtain the traveling wave profile.

\begin{remark}
The procedure to obtain the traveling wave limits and wave velocity described above can still be used for initial gas at residual saturation. In this case, the existence of traveling wave limits rely on the following limit
\begin{align*}
    \lim_{\xi\rightarrow +\infty} d_\xi \tildeSw
    = \lim_{\Sw\rightarrow 1-\Sgr} \frac{\velocity - v\,\porosity\,(1-\Sw-\Sgr)\frac{\krg+\lw\,\tildemuf}{\krg}}{\lw\,\permeability\,\dpc} = 0\,.
\end{align*}
\cite{Ashoori2011a} use a modified Brooks-Corey capillary pressure equation obtained by multiplying the original formula by the term $(1-\Sw-\Sgr)^c$, $c=0.01$. Such modification does not affect the phenomenological conclusions, but guarantees the limit above holds. In the present paper, we use the model from \cite{Ashoori2011a} with the referred modification. We stress that one can propose other modifications to the capillary pressure provided the limit above remains valid. \label{rem:limitCases}
\end{remark}

\section{Application in a linear kinetics model}
\label{sec:MRFandfw}

In this section, we apply the methodology from the previous sections to obtain traveling wave solutions for the linear kinetics foam propagation model from \cite{Ashoori2011a}.
We use them to compare the influence of the non-Newtonian apparent viscosity in the foam model.
In this model, the relative permeabilities, $\krw(\Sw)$ and $\krg(\Sw)$, are generalized Brooks-Corey expressions, fitted for the nitrogen-water flow in the Boise sandstone.
The capillary pressure has an equivalent unitary pore size distribution index, also inspired in the work of \cite{CoreyBrooks1966capillary}.
Foam generation and coalescence are controlled by expressions $\rg(\Sw)$ and $\rc(\nd)$ as follows
\begin{align}\label{eq:termsAshoori}
	\begin{aligned}
	\krw(\Sw) &:= 0.2\,\Swe^{4.2}\,,\\
	\krg(\Sw) &:= 0.94\,(1-\Swe)^{1.3}\,,\\
	\pc(\Sw) &:= 15000\,\frac{0.022\,(1-\Sw-\Sgr)^{0.01}}{\Sw-\Swc}\,,\\
	\rg(\Sw) &:= \Kc\,\nmax\ndLE(\Sw)\,,\\
	\rc(\nf) &:= \Kc\,\nmax\,\nd\,,	\\
	\ndLE(\Sw) &:= \left\{\begin{array}{ll}
		\tanh (400\,(\Sw-0.37)) & \text{ if $\Sw > 0.37$\,,}\\
		0 & \text{ if $\Sw \leq 0.37$\,,}\\
	\end{array}\right.
	\end{aligned}
\end{align}
where $\Kc$ is the kinetic generation/coalescence rate parameter.
The coefficient 15000 stems from the Leverett J-function \citep{Leverett1941} with a gas-water surface tension of 0.03 \si{N/m}. \Cref{tab:param} shows the remaining parameter values from \citep{Ashoori2011a}.
To summarize, we analyze the model described by \eqref{eq:massConsLaw}-\eqref{eq:modelFoam}, \eqref{eq:viscosityHirasaki} and \eqref{eq:termsAshoori} with parameters from \Cref{tab:param}.

\cite{Ashoori2011a} studied the incompressible foam flow considering foam as a Newtonian fluid. They used the following apparent gas viscosity formula
\begin{equation}\label{eq:viscosityAshooriTravelingWave}
	\mug^N := \mug^0\,MRF^N(\nd)\,, \qquad
	MRF^N(\nd) = 1+C_{MRF}\,\nd\,,
\end{equation}
where $C_{MRF}$ is a reference mobility reduction factor for the strongest foam.
Two well-known foam properties inspired the simplification: (1) foamed gas mobility reduction is large and nearly constant at high water saturation, and (2) there exists a limiting capillary pressure, where foam collapses abruptly.
Parameter $\alpha$ is the one that minimizes the mean squared difference between the gas mobilities under equilibrium conditions, i.e.,
\begin{align}\label{eq:alpha}
	\min_{\alpha} \int_{\Swc}^{1-\Sgr} \left(\frac{\krg(\Sw)}{\mug^{LE}(\Sw,\ndLE(\Sw),\velocity)}-\frac{\krg(\Sw)}{\mug^0\,MRF^N(\ndLE(\Sw))}\right)^2\, d\Sw\,.
\end{align}
for $\velocity = \num{ 29.3 } \si{\mu m/s }$. This mapping generalizes the one proposed by \cite{Zavala2021}.

\begin{table}[htb]
\centering
\caption{Parameters for nitrogen and water flow in a Boise sandstone \citep{Ashoori2011a}.}
\label{tab:param}
\begin{tabular}{lll}
	\hline
	Property & Symbol & Value\\ [3pt] \hline
	Newtonian gas viscosity coefficient & $C_{MRF}$ & \num{18500} [-]\\
	Permeability & $ \permeability $ & \num{e-12} \si{ m^2 }\\
	Kinetic parameter & $ \Kc $ & 200; 1; 0.01  \si{ s^{-1} }\\
	Reference foam texture & $ \nmax $ & \num{ 8e13 } \si{ m^{-3} } \\
	Residual gas saturation & $ \Sgr $ & \num{ 0.18 } [-] \\
	Residual (connate) water saturation & $ \Swc $ & \num{ 0.2 } [-] \\
	Total superficial velocity & $\velocity$ & \num{ 2.93e-5 } \si{m/s } \\
	Non-Newtonian gas viscosity coefficient & $\alpha$ & \num{ 2.44e-16 } \si{ Pa.s^{\frac23}.m^{\frac{10}3} } \\
	Water viscosity & $ \muw $ & \num{ e-3 } \si{ Pa.s }\\
	Gas viscosity in the absence of foam & $ \mug^0 $ & \num{ 2e-05 } \si{ Pa.s }\\
	Porosity & $ \porosity $ & \num{ 0.25 } [-]
	\\ \hline
\end{tabular}
\end{table}

In what follows, we compute traveling wave profiles for the linear kinetic model using the non-Newtonian apparent gas viscosity \cref{eq:viscosityHirasaki}. We compare those profiles with the ones obtained using the Newtonian apparent viscosity \cref{eq:viscosityAshooriTravelingWave}.
We analyze two scenarios. The first one reproduces the example found in \citep{Ashoori2011a}. The second one show a case where the non-Newtonian foam model brings significant information.
We also show how apparent total viscosity changes with flow velocity in the equilibria, which helps understanding the connection between Newtonian and non-Newtonian foam models.
Hereafter, the superscript `$N$' identifies the properties of the Newtonian and `$S$' the properties of the non-Newtonian models.
\Cref{tab:travelingWaveStates} shows the equilibria computed using \crefrange{eq:travelingODElimitSystem_fgm}{eq:travelingODElimitSystem_rgrcp} for the Newtonian and non-Newtonian models in the two scenarios.

\sisetup{round-mode=figures,round-precision=4,scientific-notation=true}
\begin{table}[htb]
\centering
\caption{Parameter values at limit states and traveling wave velocities in two scenarios used to compare Newtonian and non-Newtonian models. We use $\velocity = 29.3 \si{ \mu m/s }$.}
\label{tab:travelingWaveStates}
\begin{tabular}{l|ll|ll}
    \hline
	& \multicolumn{2}{c|}{Scenario 1} & \multicolumn{2}{c}{Scenario 2} \\
	& \multicolumn{2}{c|}{$f_w^-$ = 0.268; $S_w^+$ = 0.72} & \multicolumn{2}{c}{$f_w^-$ = 0.9; $S_w^+$ = 0.819} \\ [3pt] \hline
	Parameter & Newtonian & non-Newt. & Newtonian & non-Newt.\\\hline
	$\Swm$ [-] & \num{0.372004} & \num{0.371940} & \num{0.501128} & \num{0.467238}\\
	$\ndm$ [-] & \num{0.664985} & \num{0.650301} & \num{1.0} & \num{1.0}\\
	$\ndp$ [-] & \num{1.0} & \num{1.0} & \num{1.0} & \num{1.0}\\
	$\ug^+$ [\si{\mu m/s}] & \num{7.250802325588347e-02} & \num{6.11115662115901e-03} & \num{8.781610e-05} & \num{3.192801e-07}\\
	$\vert \nabla \pressure \vert^-$ [\si{MPa/m}] & \num{8.565506} & \num{8.579051} & \num{2.737615} & \num{4.520126}\\
	$\vert \nabla \pressure \vert^+$ [\si{MPa/m}] & \num{3.059092e-01} & \num{3.066042e-01} & \num{1.474962e-01} & \num{1.474966e-01}\\
	$v$ [\si{m/s}] & \num{2.456937e-04} & \num{2.464111e-04} & \num{3.686905e-05} & \num{3.331801e-05}
    \\ \hline
\end{tabular}
\end{table}
\sisetup{round-mode=off,scientific-notation=false}

\subsection{Scenario 1}
\label{sec:scenario1_sim}

Scenario 1 corresponds to the flow data from the literature \citep{Ashoori2011a} with the limit states $\Swp = 0.72$ and $\fwm = 0.268$.
The autonomous system \eqref{eq:travelingODE2} can be rewritten in the standard form:
\begin{align}\label{eq:travelingODE3}
	d_\xi\left[
 	\begin{array}{c}
 	    \tildeSw \\ \tildenD
 	\end{array}
 	\right]
 	= F(\tildeSw,\tildenD); & \quad \text{where} \nonumber\\
 	F(\Sw,\nd)
 	&:= 
 	\left[\begin{array}{c}
 		\dfrac{\left(\ug^+ - \frac{\krg\,\velocity}{\krg+\lw\,\mu}\right) - v\,\porosity\,(\Sw-\Swp)}{-\frac{\krg}{\krg+\lw\,\mu}\,\lw\,\permeability\,\dpc}\\
 		\\
 		\dfrac{\porosity\,(1-\Sw)\,R}{\nmax\,(\ug^+-v\,\porosity\,(1-\Swp))}
 	\end{array}\right] \nonumber\\
    \text{and}\quad \mu &:= \mug^0 + \frac{\alpha\,\nmax}{v^{\frac13}}\,\left(\frac{1-\Sw}{\big\vert \Sw^{crit} - \Sw\big\vert }\right)^{\frac13}\,\nd,
\end{align}
One standard approach for proving the existence of the solution connecting two equilibria in an autonomous system is based on the analysis of the ODE's phase space, as done by \cite{Lozano2021}. We repeat some steps of their analysis.

We start by proving that:
\begin{enumerate}
\item $F(\Sw,\nd) = 0$ if and only if $\nd = \ndLE(\Sw)$ and $\ug^+ - \krg/(\krg+\lw\,\mu)\,\velocity = v\,\porosity\,(\Sw-\Swp)$;
\item The curve $\ug^+ - \krg\,\velocity/(\krg+\lw\,\mu)$ restricted to $\nd = \ndLE(\Sw)$ intersects the straight line $v\,\porosity\,(\Sw-\Swp)$ in at most two values of $\Sw \in (\Swc,1-\Sgr)$. 
\end{enumerate}
Since $F(\Sw^+,\nd^+) = F(\Sw^-,\nd^-) = 0$, those are the only two roots of $F$ in $(\Swc,1-\Sgr) \times [0,1]$.

Second, let us comment on the dynamics in the neighborhood of the limit states of Scenario 1 in \cref{tab:travelingWaveStates}.
In two-dimensional phase space, this analysis reduces to look at the real part's sign of the eigenvalues of the Jacobian matrix $(\partial_{\Sw} F, \partial_{\nd} F)$;
positive signs indicate source directions, and negative signs indicate sink directions.
Since $F(\Sw,\nd) = 0$ only over the equilibria, the orbit (solution of \cref{eq:travelingODE3}) exits if there is at least one source direction in the left state and at least one sink direction in the right state.
\Cref{tab:eigenval} shows the eigenvalues for states $-$ and $+$ using the non-Newtonian model.
In all cases, the states $-$ have eigenvalues with positive real parts (sources), and states $+$ have eigenvalues' real parts with opposite signs (saddle points), which proves existence of an orbit, i.e., traveling wave solution.

\sisetup{round-mode=figures,round-precision=4,scientific-notation=true}
\begin{table}[htb]
 	\centering
 	\caption{Eigenvalues of the Jacobian matrix 
 	$(\partial_{\Sw} F, \partial_{\nd} F)$ at the equilibrium states $-$ and $+$ of Scenario 1. The classification follows a typical nomenclature \citep{Guckenheimer1984}.}
 	\label{tab:eigenval}
 	\begin{tabular}{ccccc}
    \hline
 	$\Kc$ [\si{1/s}] & State & $\lambda_1$ [\si{1/m}] & $\lambda_2$ [\si{1/m}] & Classification\\ \hline
 	200 & $-$ & \num{ 108891.08524704888 } & \num{ 1715138.3958535134 } & Source\\
 	200 & $+$ & \num{ -1019924.1847877622 } & \num{ 809184.9862640935 } & Saddle\\
 	1 & $-$ & \num{ 1091+2854i } & \num{ 1091-2854i } & Spiral source\\
 	1 & $+$ & \num{ -1019924.1847877622 } & \num{ 4045.924931320468 } & Saddle\\
 	0.01 & $-$ & \num{ 12087.06493922236 } & \num{ 772.5749890998247 } & Source\\
 	0.01 & $+$ & \num{ -1019924.1847877622 } & \num{ 40.45924931320468 } & Saddle
    \\ \hline
 	\end{tabular}
\end{table}
\sisetup{round-mode=off,scientific-notation=false}

\begin{figure}[htb]
\centering
	\includegraphics[width=.48\linewidth]{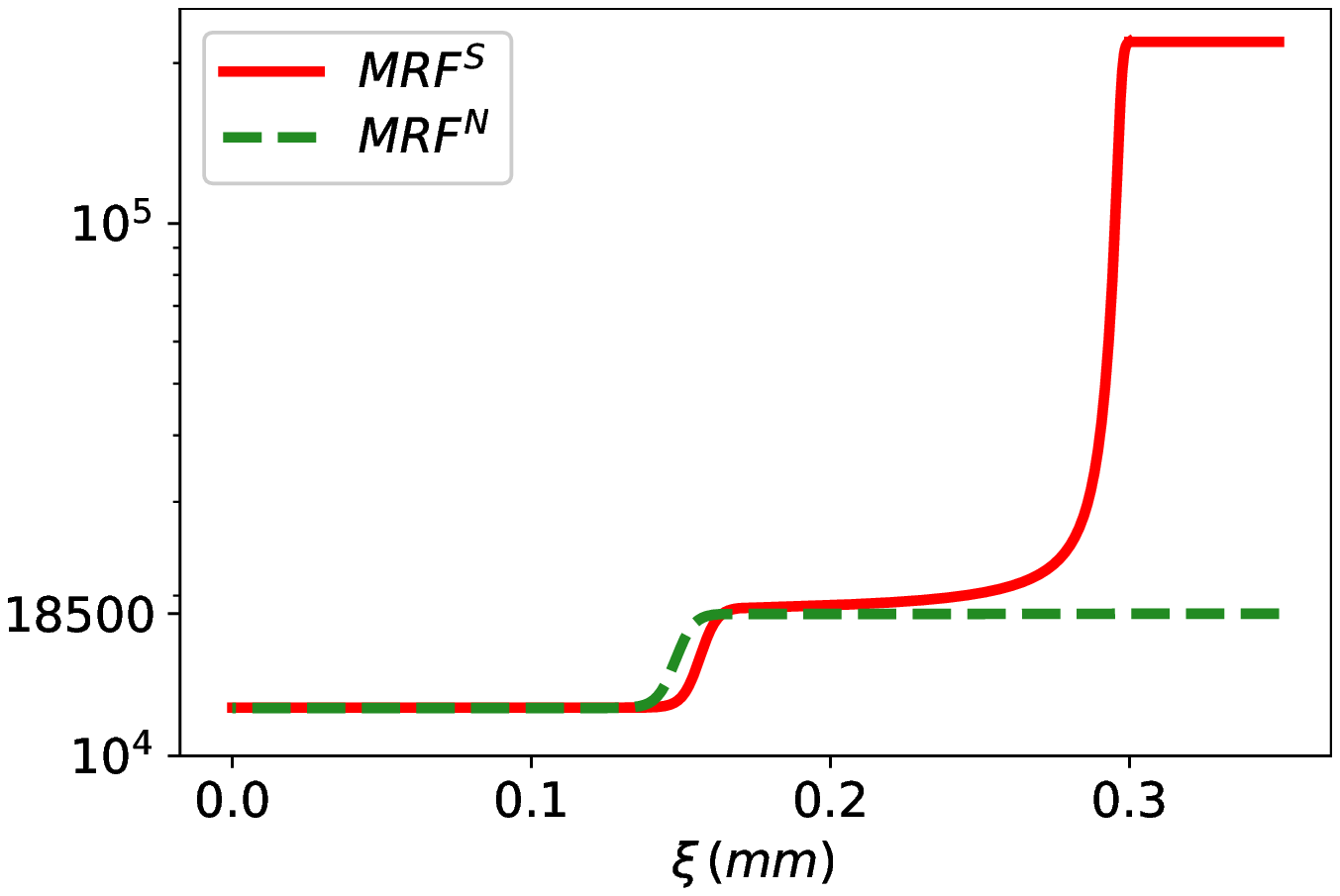}
	\hfill
	\includegraphics[width=.48\linewidth]{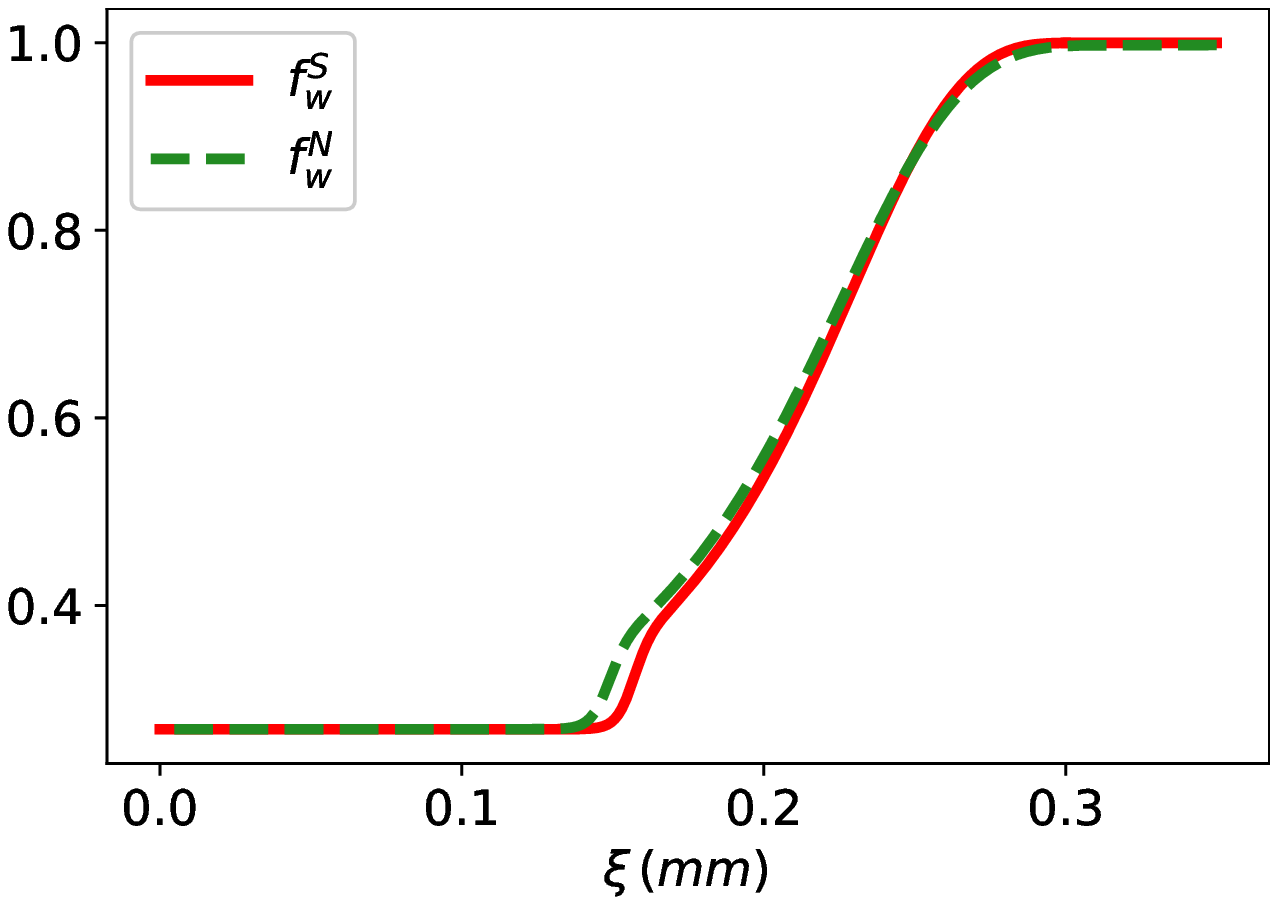}
	\caption{Mobility Reduction Factor and $f_w$ profiles in the non-Newtonian (filled lines) and Newtonian (dashed lines) models using $\Kc=200$ \si{s^{-1}} in Scenario 1.}
	\label{fig:orbitsAndWaveProfilesMRF_scenario1}
\end{figure}

\Cref{tab:travelingWaveStates} shows that the relative difference for the states $\ndm$ is lower than 3\%, and the values of $\ug^+$ are approximately zero. The remaining parameter values vary less than 1\% between the models. Note that the Newtonian model's values in scenario 1 coincide with those from \citep{Ashoori2011a}.
The solution profiles for $\Sw$ and $\nd$ for the non-Newtonian case are very similar to the ones of the Newtonian model found in \citep{Ashoori2011a}.
\Cref{fig:orbitsAndWaveProfilesMRF_scenario1} shows the MRF and $f_w$ profiles.
Although the MRF we observe in the left panel of \cref{fig:orbitsAndWaveProfilesMRF_scenario1} a significant qualitative and quantitative change in MRF, this does not affect fractional flux curves as depicted in the right panel of the same figure.

\subsection{Scenario 2}

The second scenario explores higher differences between the Newtonian and non-Newtonian behaviors by taking a higher initial water saturation, $\Swp = 0.819$, and injecting a mixture with lower foam quality, $\fwm = 0.9$.
This scenario, usually denoted by the term low-quality regime, was not based on experimental data from the literature; it was built to enlarge the differences between Newtonian and non-Newtonian regimes describing foam flow in porous media. Scenario 2 addresses a fundamental scientific discussion on understanding the non-Newtonian nature of the foam displacement.
The analysis of the existence of traveling wave solutions for Scenario 2 is analogous to the one from \cref{sec:scenario1_sim} and, therefore, we omit it.

\begin{figure}[htb]
\centering
	\includegraphics[width=.48\linewidth]{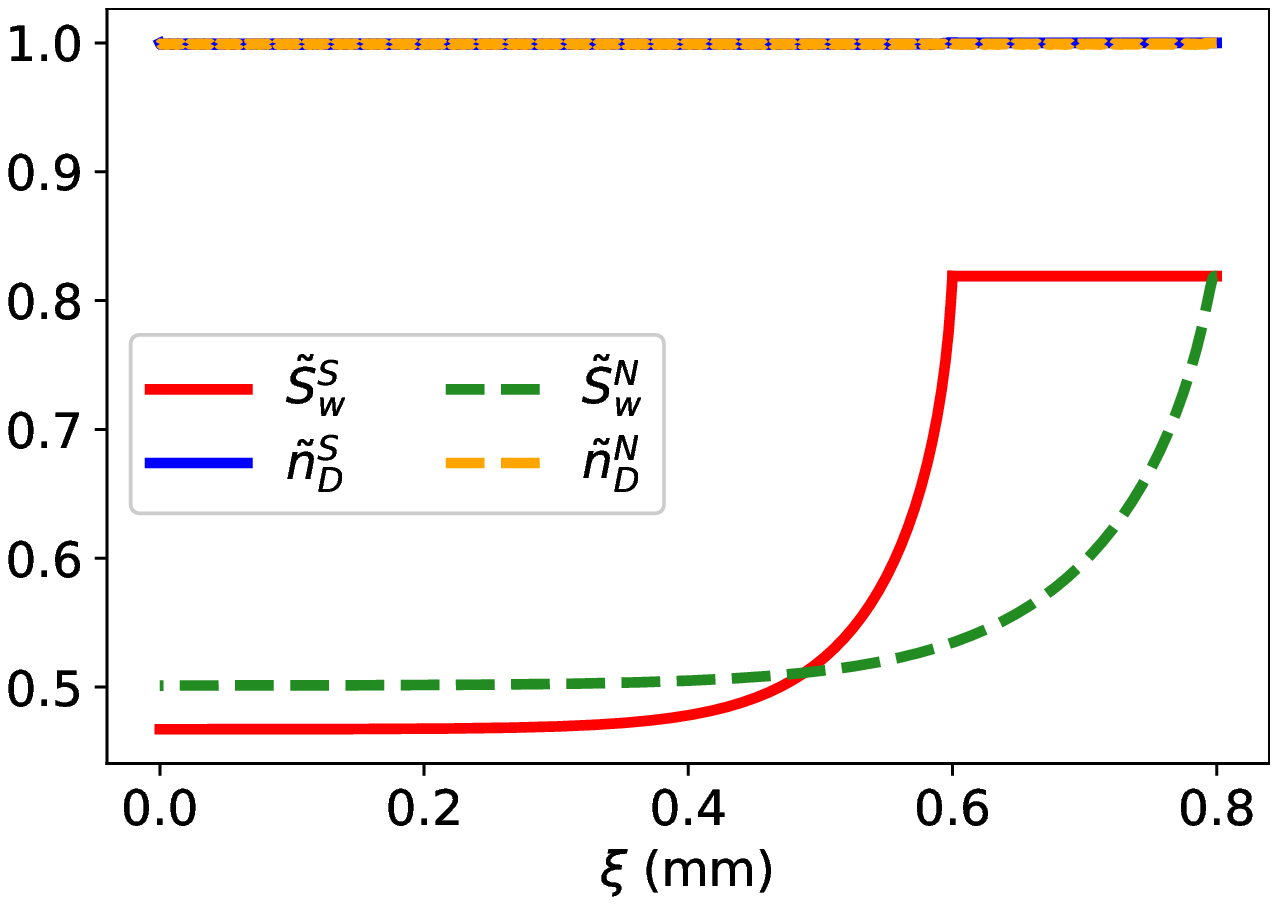}
	\hfill
	\includegraphics[width=.48\linewidth]{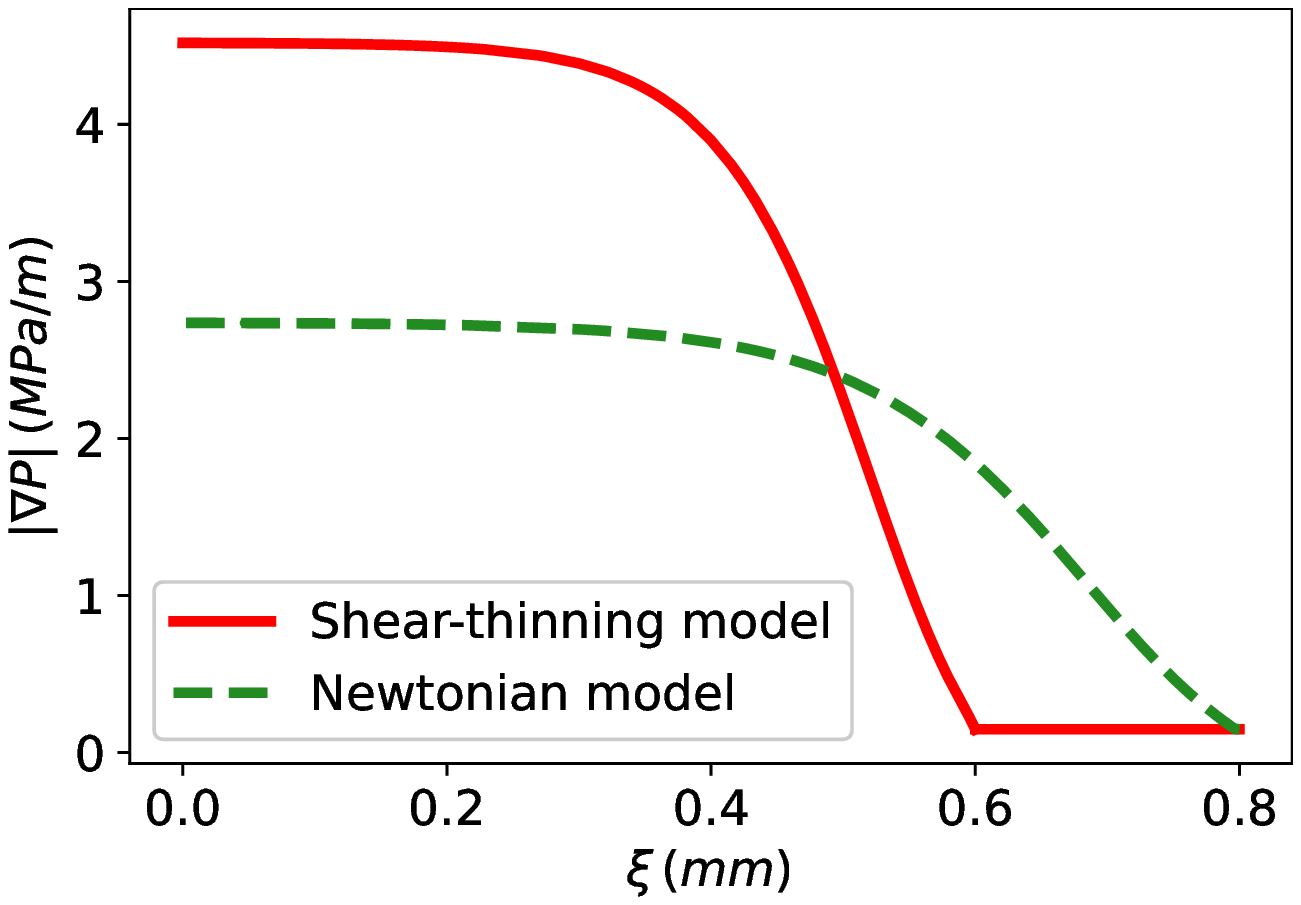}
	\caption{Wave profiles in the non-Newtonian (filled lines) and Newtonian (dashed lines) models using $\Kc=200$ \si{s^{-1}} in Scenario 2. Left: $\tildeSw(\xi)$ and $\tildenD(\xi)$. Right: pressure gradient obtained using the formulas from \cref{sec:model}.}
	\label{fig:orbitsAndWaveProfilesLowQuality}
\end{figure}

\begin{figure}[htb]
\centering
	\includegraphics[width=.48\linewidth]{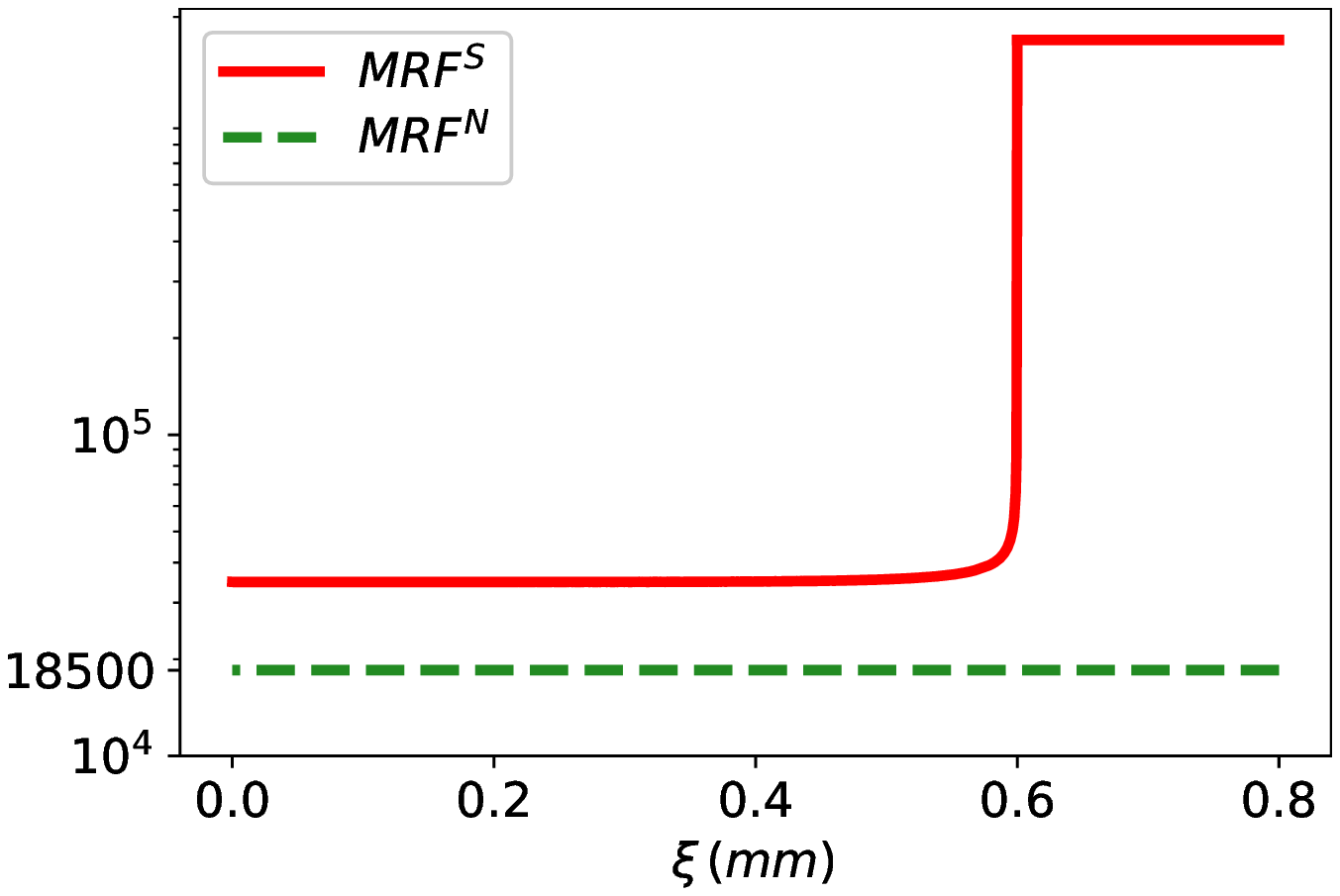}
	\hfill
	\includegraphics[width=.48\linewidth]{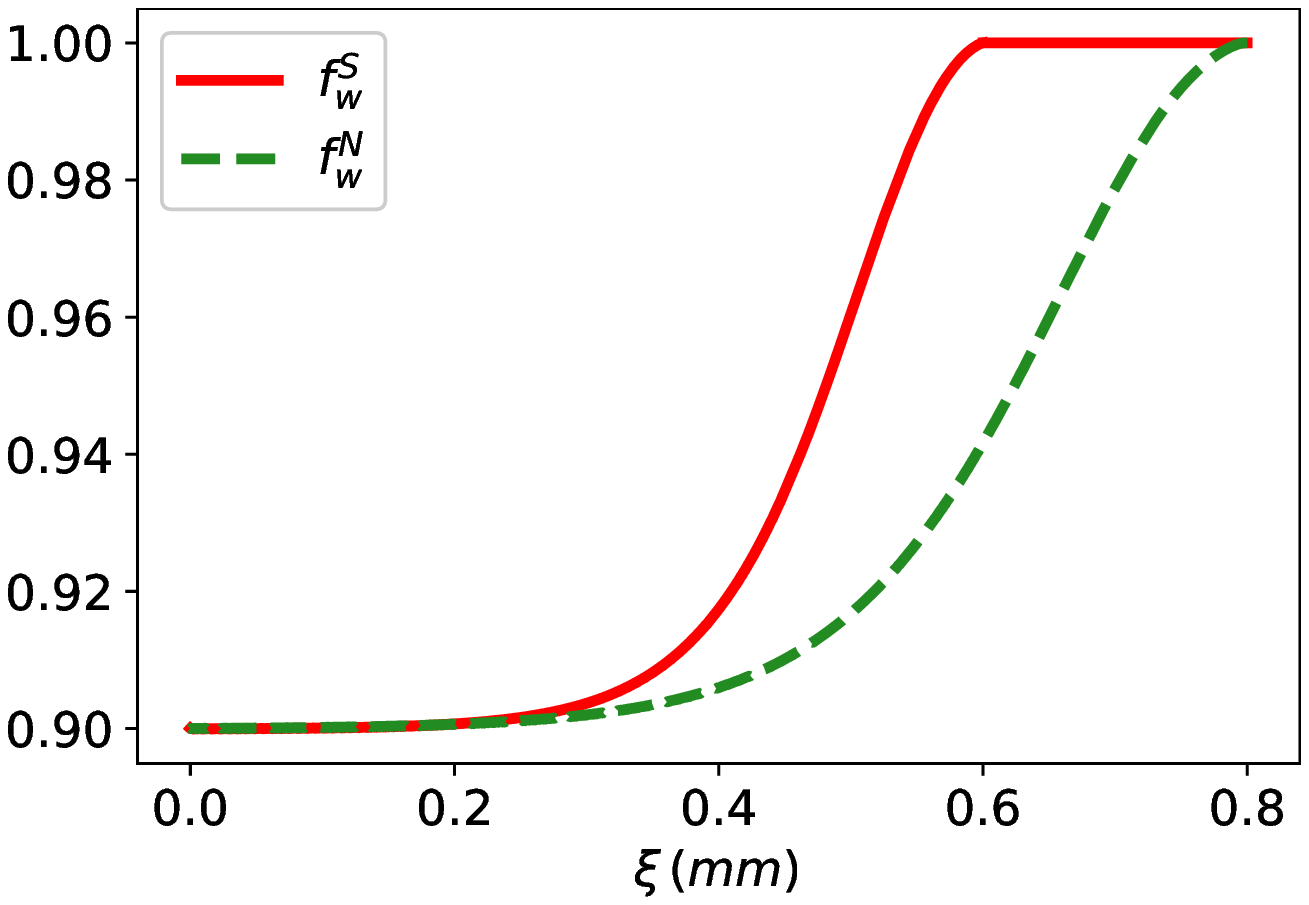}
	\caption{Mobility Reduction Factor and $f_w$ profiles in the non-Newtonian (filled lines) and Newtonian (dashed lines) models using $\Kc=200$ \si{s^{-1}} in Scenario 2.}
	\label{fig:orbitsAndWaveProfilesMRF_scenario2}
\end{figure}

\Cref{tab:travelingWaveStates} shows the equilibria for the Newtonian and non-Newtonian models in Scenario 2.
We note that the Newtonian traveling wave velocity $v$, water saturation $\Swm$, and pressure gradient $\vert \nabla\pressure\vert^-$ differ from the corresponding values of the non-Newtonian values 11\%, 7\%, and 39\%, respectively.
\Cref{fig:orbitsAndWaveProfilesLowQuality} shows the traveling wave profiles of $\tildeSw$ and $\tildenD$, and the pressure gradient over the traveling wave for the non-Newtonian foam model using $\Kc=200$ \si{s^{-1}}. Besides the difference in the limit states, the non-Newtonian foam model profiles are sharper than the ones in the Newtonian model.
\Cref{fig:orbitsAndWaveProfilesMRF_scenario2} shows the MRF and $f_w$ over the traveling wave. We use \cref{eq:viscosityHirasakiTravelingWave} to compute $MRF^S$. As in scenario 1, MRF is qualitatively and quantitatively different between the models. In this case, however, the fractional flux profile changes between the models leading to different water saturation profiles, see left panel of \cref{fig:orbitsAndWaveProfilesLowQuality}.

\subsection{Traveling wave equilibria for different flow velocities}

\begin{figure}[htb]
	\centering
	\includegraphics[width=.48\linewidth]{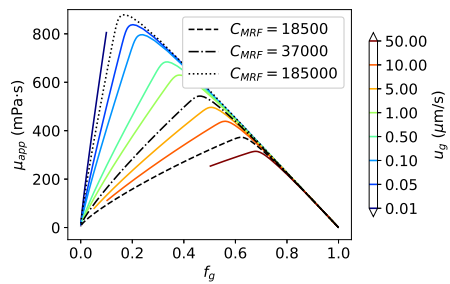}
	\caption{Apparent viscosity $\mu_{app}$ as a function of $\ug$ and $\fg$, all using LE foam, for the non-Newtonian foam model. The dashed and dotted lines represent Newtonian total apparent viscosity with a given $C_{MRF}$.}
	\label{fig:muApp}
\end{figure}

\Cref{fig:muApp} shows isolines of the superficial gas velocity $\ug$ as a function of the foam quality $\fg$ and the total apparent viscosity $\mu_{app}$ defined in \eqref{eq:muApp}.
All quantities are taken considering LE conditions and $\velocity \in [0.1\si{\mu m/s},100\si{\mu m/s}]$, a range that covers laboratory and field experimental values.
One can see that the lower is the superficial gas velocity the stronger is the foam (higher $\mu_{app}$).
In agreement with \citep{Eide2020,Gassara2020}, the high-quality regime region (after the peak of the strongest foam) shrinks when the superficial gas velocity increases.
In the Newtonian model, total apparent viscosity and fractional fluxes depend only on $\Sw$, so there is a single curve for each $C_{MRF}$. We plot the curves $C_{MRF} = \num{37000}$ and $C_{MRF} = \num{185000}$ in \cref{fig:muApp} to show that one should vary $C_{MRF}$ so that the Newtonian model better represent the non-Newtonian gas viscosity accordingly to the flow regime.

\begin{remark}
	Scenario 2 evidences some difference between the traveling wave solutions of non-Newtonian and Newtonian models. The parameter values in this scenario assume the injection of a mixture with low gas concentration. Although it is uncommon, such configuration occurs in industrial application, e.g., surfactant alternated gas (SAG) injection in the Enhanced Oil Recovery (EOR). During this procedure, pure water (containing surfactant) and pure gas are injected into the reservoir periodically resulting in alternated low and high gas concentrations inside the porous medium.
\end{remark}

\begin{remark}\label{rem:odeSolver}
We used the explicit Runge-Kutta method RK5(4) \citep{Shampine1986} to solve \eqref{eq:travelingODE2} in the variable $\eta := -\xi$ on $(0,\eta_F)$ with initial conditions $(\tildeSw,\tildenD)(0) = (\Swp-10^{-6},\ndp)$ and time steps bounded by $\eta_F/1000$.
This change of variables, $\xi$ to $-\eta$, helps the numerical solver to find the correct solution.
\end{remark}

\section{Discussion and Conclusions}
\label{sec:conclusions}

In this paper, we detailed the procedure of obtaining traveling wave solutions for a range of models describing propagation of foam in porous media. For that, we reformulate the Hirasaki and Lawson's apparent viscosity in terms of the foam model's unknowns $\Sw$, $\nd$ and $\velocity$. We restrict the analysis to the case of constant total superficial velocity $\velocity$ as in previous works. It is worth mentioning that the reformulated apparent viscosity formula can be used in other contexts. One example is to reconstruct fractional fluxes directly from the experimental data.

In order to evidence the potential of the proposed methodology, we find the traveling wave solutions for the linear kinetic model from \cite{Ashoori2011a}. We use the solutions to quantify the importance of considering the non-Newtonian apparent viscosity in the model. We proposed a mapping procedure to map one model in the other. This procedure involves the solution of a minimization problem for the mean square difference between gas mobilities in the two models. We conclude that:
\begin{itemize}
	\item For the cases reproduced from \cite{Ashoori2011a}, the non-Newtonian viscosity brings no new relevant information. This is expected in such a high-quality regime, as observed in several other works, e.g., \citep{UT,castillo2020fractional,valdez2021foam}.
    \item In the low-quality example, we identify a foam flow regime with stronger non-Newtonian fluid behavior, and obtain more significant discrepancies between the two models. Our example showed a relative difference of 39\% in the pressure gradient and 11\% in the traveling wave velocity. Note that estimating the traveling wave velocity is crucial to predict the gas breakthrough during an enhanced recovery process in porous media. It is worth mentioning that we achieve such contrast even using the same constant total superficial velocity from Scenario 1.
    \item The mobility reduction factor gives significantly less information on the foam flow than the fractional flux curves. In the two scenarios we analyze, the mobility reduction factors between Newtonian and non-Newtonian presented qualitative and quantitative differences. However, the water saturation and foam texture wave profiles coincide qualitatively in the first scenario, while, in the second one, they don't.
    \item For different total flow velocities and quality regimes, one may still use the Newtonian gas viscosity formula, although the parameters should be adjusted so that the shear-thinning behavior of foam can be accurately represented.
\end{itemize}

\bibliography{references_abbrev}

\end{document}